\documentclass[lettersize,journal]{IEEEtran}
\usepackage{amsmath,amsfonts}
\usepackage{algorithmic}
\usepackage{algorithm}
\usepackage{array}
\usepackage[caption=false,font=normalsize,labelfont=sf,textfont=sf]{subfig}
\usepackage{textcomp}
\usepackage{stfloats}
\usepackage{url}
\usepackage{verbatim}
\usepackage{graphicx}
\usepackage{cite}
\usepackage{booktabs}
\usepackage{multirow}
\usepackage{multicol}
\usepackage[nohyperlinks,nolist]{acronym}
\usepackage{siunitx}
\usepackage[para]{threeparttable}

\usepackage{eso-pic}

\newcommand{\positformat}[2]{posit$\langle #1, #2 \rangle$}

\begin{document}

\title{Increasing the Energy Efficiency of Wearables Using Low-Precision Posit Arithmetic with PHEE}

\author{David~Mallasén,
        Pasquale~Davide~Schiavone,~\IEEEmembership{Member,~IEEE},
        Alberto~A.~Del~Barrio,~\IEEEmembership{Senior~Member,~IEEE},
        Manuel~Prieto-Matias,
        and~David~Atienza,~\IEEEmembership{Fellow,~IEEE}
\thanks{D.~Mallasén, P.~D.~Schiavone, and D.~Atienza are with the Embedded Systems Laboratory (ESL), École Polytechnique Fédérale de Lausanne (EPFL), 1015 Lausanne, Switzerland. E-mails: \{david.mallasen, davide.schiavone, david.atienza\}@epfl.ch}
\thanks{A.~A.~Del~Barrio, and M.~Prieto-Matias are with the Facultad de Informática, Universidad Complutense de Madrid, 28040 Madrid, Spain. E-mails: \{abarriog, mpmatias\}@ucm.es}
\thanks{Manuscript received -; revised -.}}

\AddToShipoutPictureBG*{%
  \AtPageUpperLeft{%
    \setlength\unitlength{1in}%
    \hspace*{\dimexpr0.5\paperwidth\relax}
    \makebox(0,-0.5)[c]{\footnotesize This article has been accepted for publication in a future issue of this journal, but has not been fully edited. Content may change prior to final publication.}%
}}
\AddToShipoutPictureBG*{%
  \AtPageUpperLeft{%
    \setlength\unitlength{1in}%
    \hspace*{\dimexpr0.5\paperwidth\relax}
    \makebox(0,-0.8)[c]{\footnotesize Citation information: DOI 10.1109/TCASAI.2026.3652126, IEEE Transactions on Circuits and Systems for Artificial Intelligence}%
}}
\AddToShipoutPictureBG*{%
  \AtPageLowerLeft{%
    \setlength\unitlength{1in}%
    \hspace*{\dimexpr0.5\paperwidth\relax}
    \makebox(0,0.8)[c]{\footnotesize © 2026 IEEE. Personal use of this material is permitted. Permission from IEEE must be obtained for all other uses, in any current}%
}}
\AddToShipoutPictureBG*{%
  \AtPageLowerLeft{%
    \setlength\unitlength{1in}%
    \hspace*{\dimexpr0.5\paperwidth\relax}
    \makebox(0,0.6)[c]{\footnotesize or future media, including reprinting/republishing this material for advertising or promotional purposes, creating new collective}%
}}
\AddToShipoutPictureBG*{%
  \AtPageLowerLeft{%
    \setlength\unitlength{1in}%
    \hspace*{\dimexpr0.5\paperwidth\relax}
    \makebox(0,0.4)[c]{\footnotesize works, for resale or redistribution to servers or lists, or reuse of any copyrighted component of this work in other works.}%
}}

\begin{acronym}[]
    \acro{ALU}{arithmetic logic unit}
    \acro{ASIC}{application-specific integrated circuit}
    \acro{AUC}{area under the ROC curve}
    \acro{AI}{artificial intelligence}
    \acro{BiCG}{biconjugate gradient}
    \acro{CISC}{complex instruction set computer}
    \acro{CNN}{convolutional neural network}
    \acro{CG}{conjugate gradient}
    \acro{CV-X-IF}{Core-V eXtension Interface}
    \acro{DNN}{deep neural network}
    \acro{ECG}{electrocardiogram}
    \acro{FPU}{floating-point unit}
    \acro{FMA}{fused multiply-add}
    \acro{FPGA}{field-programmable gate array}
    \acro{FF}{flip-flop}
    \acro{FFT}{fast Fourier transform}
    \acro{FPR}{false positive rate}
    \acro{GPU}{graphics processing unit}
    \acro{GAN}{generative adversarial networks}
    \acro{GEMM}{general matrix multiplication}
    \acro{HPC}{high-performance computing}
    \acro{HLS}{high-level synthesis}
    \acro{ISA}{instruction set architecture}
    \acro{IP}{intellectual property}
    \acro{IMU}{inertial measurement unit}
    \acro{LUT}{lookup table}
    \acro{LSB}{least significant bit}
    \acro{ML}{machine learning}
    \acro{MAC}{multiply-accumulate}
    \acro{MSE}{mean squared error}
    \acro{MaxAbsE}{maximum absolute error}
    \acro{MFCC}{mel-frequency cepstral coefficient}
    \acro{NaN}{Not a Number}
    \acro{NaR}{Not a Real}
    \acro{NN}{neural network}
    \acro{OS}{operating system}
    \acro{PAU}{posit arithmetic unit}
    \acro{PRAU}{Posit and quiRe Arithmetic Unit}
    \acro{RISC}{reduced instruction set computer}
    \acro{RTL}{register-transfer level}
    \acro{RMSE}{root mean squared error}
    \acro{ROC}{receiver operating characteristic}
    \acro{SIMD}{single instruction, multiple data}
    \acro{SoC}{system-on-chip}
    \acro{TPR}{true positive rate}

\end{acronym}

\maketitle

\begin{abstract}
Wearable edge AI biomedical devices are increasingly being used for continuous patient health monitoring, enabling real-time insights and extended data collection without the need for prolonged hospital stays. These devices must be energy efficient to minimize battery size, improve comfort, and reduce recharging intervals. This paper investigates the use of specialized low-precision arithmetic formats to enhance the energy efficiency of edge AI biomedical wearables. Specifically, we explore posit arithmetic, a floating-point-like representation, in two biomedical applications that leverage supervised and unsupervised learning algorithms: cough detection for chronic cough monitoring and R peak detection in ECG analysis. Our results reveal that 16-bit posits can replace 32-bit IEEE 754 floating point numbers with minimal accuracy loss in cough detection. For R peak detection, posit arithmetic achieves satisfactory accuracy with as few as 10 or 8 bits, compared to the 16-bit requirement for floating-point formats.

To validate these findings beyond algorithm-level simulations, we introduce PHEE, a modular and extensible architecture that integrates the Coprosit posit coprocessor within a RISC\nobreakdash-V-based system. Using the X\nobreakdash-HEEP framework, PHEE serves as a proof-of-concept platform to quantify the practical energy benefits of low-precision posits in edge AI systems. Post-synthesis results targeting 16~nm TSMC technology show that the posit hardware targeting these ML-based biomedical applications can be 38\% smaller and consume up to 42.3\% less power at the functional unit level, with no performance compromise. These findings establish the potential of low-precision posit arithmetic to significantly improve the energy efficiency of edge AI biomedical devices.
\end{abstract}


\section{Introduction} \label{sec:introduction}
\IEEEPARstart{W}{earable} devices are widely used in medical contexts to enable continuous monitoring of patient health using \ac{ML}-based applications~\cite{wei2020Review}. These devices collect data through sensors distributed throughout the body, process it locally, and store or transmit it externally. This approach allows physicians to collect patient data over long periods without requiring hospital stays. In addition, edge \ac{AI} wearable devices provide patients with real-time health insights, enabling them to take preventive measures. Since these devices are worn on the body, they must be battery-powered, comfortable, and require minimal user interaction~\cite{najafi2024VersaSens}. Consequently, their energy efficiency is of utmost importance in minimizing battery size and extending recharging intervals~\cite{liu2023UltraLow}.

Some researchers have tackled the challenge of energy consumption by using energy harvesting technologies, which include sources such as solar or thermoelectric power, radio frequency, and human energy converted into chemical, thermoelectric, or kinetic energy~\cite{chong2019Energy}. Others have focused on iteratively extending the feature sets used by detection or classification algorithms until a specified confidence threshold is reached~\cite{ferretti2022INCLASS,wang2024ACE,momeni2022CAFS}. This approach reduces the energy cost of running complex algorithms that depend on extracting numerous feature sets from input sensors. A related strategy involves combining lightweight algorithms, which can operate efficiently in most scenarios, with more complex and accurate algorithms that are activated only when necessary~\cite{degiovanni2023Adaptive}.

Unfortunately, little attention has been paid to the impact of arithmetic representations utilized in non-\ac{NN} algorithms running on wearable devices. Typically, \ac{ML}-based algorithms operate using integer or fixed point formats~\cite{dekimpe2022ECG}, and if the application requires a higher dynamic range or precision, it is switched to using 32-bit floating point numbers. However, a considerable gap exists between these options, especially when the required dynamic range cannot be easily met by fixed-point numbers, and energy consumption is closely linked to the bit-width of the data. For example, a 32-bit floating point addition can consume more than twice the energy compared to a 16-bit addition, and this disparity increases to more than three times for multiplication in a 45~nm technology node~\cite{horowitz201411}. Using low-precision formats presents an additional solution to enhance the energy efficiency of these \ac{ML} applications. If a 16-bit floating-point representation provides the necessary accuracy for a given application, its lower energy cost makes it the ideal candidate to replace single-precision numbers~\cite{haidar2018Design}.

This paper explores the use of specialized arithmetic formats in \ac{ML}-based biomedical applications designed for edge \ac{AI} wearables. In particular, we focus on the detection of cough for the continuous monitoring of patients with chronic cough, as well as R peak detection in \ac{ECG} analysis. These applications rely on the random forest classifier and the k-means clustering algorithms, classic examples of supervised and unsupervised learning. We explore the use of posit arithmetic~\cite{positworkinggroup2022Standard}, a floating point-like representation that has proven useful in domains such as signal processing for radio astronomy~\cite{gunaratne2023Evaluation}, scientific computing~\cite{chien2020Posit,mallasen2024BigPERCIVAL}, and \acp{NN}~\cite{raposo2021Positnn,murillo2020Deep}. Unlike prior work that primarily explored FP16 or FP32 formats, we target low-precision posits and demonstrate that they can maintain high accuracy in biomedical ML workloads, outperforming FP8 formats. This precision reduction is critical for wearables, where energy and memory constraints dominate design decisions.

To reinforce these findings through ASIC-level energy and area analysis, we present Posit enabled x-HEEp (PHEE), our open-source hardware implementation that integrates a posit arithmetic coprocessor into a microcontroller-level RISC\nobreakdash-V \ac{SoC}. PHEE features Coprosit, a posit arithmetic coprocessor compatible with the \ac{CV-X-IF} interface, connected to the cv32e40px RISC\nobreakdash-V core of the X\nobreakdash-HEEP platform~\cite{machetti2024XHEEP}. This setup allows us to leverage the results achieved in arithmetic simulations and to bridge algorithm-level precision optimization with hardware-level evaluation. Using PHEE, we provide post-synthesis \ac{ASIC} results in 16~nm technology, quantifying area and energy consumption of the dedicated posit hardware unit in PHEE, enabling a comparison with the alternative method of using standard floating-point numbers.

In summary, our main contributions are as follows:
\begin{itemize}
    \item We compare the use of 16-bit and 32-bit arithmetic representations in a cough detection application designed for the continuous monitoring of chronic cough patients. The results indicate that 16-bit posits achieve comparable accuracy to 32-bit floats, while bfloat16 and half-precision floats show a more significant drop in accuracy.
    \item In the context of R peak detection in \acp{ECG}, we analyze the effectiveness of low-precision representations and find that posits with 10 or even 8 bits maintain an F1 score above 0.9. In contrast, at least 16 bits are required in the floating-point case to obtain a comparable accuracy.
    \item To move beyond theoretical simulations and assess hardware feasibility, we present PHEE. This is an open-source RISC\nobreakdash-V \ac{SoC} integrating Coprosit, our standalone posit coprocessor. This platform enables the native execution of various posit sizes, similar to the F and D standard RISC\nobreakdash-V extensions, by leveraging the custom Xposit extension presented in~\cite{mallasen2022PERCIVAL}.
    \item \ac{ASIC} synthesis results for TSMC 16~nm technology highlight the area and energy savings associated with utilizing low-bit-width posit hardware compared to standard IEEE floating-point representation. This is studied in detail for the \ac{FFT}, which is the most computationally intense kernel present in the cough detection application.
\end{itemize}

The rest of this paper is organized as follows: Section~\ref{sec:background} introduces posit arithmetic and the X\nobreakdash-HEEP hardware platform. Section~\ref{sec:related_work} reviews related work. Section~\ref{sec:biomed_apps} describes the biomedical applications we have studied in this work and discusses the results obtained by the different arithmetic formats. Section~\ref{sec:PHEE} outlines PHEE's architecture, including that of the Coprosit posit arithmetic coprocessor. Section~\ref{sec:asic_results} discusses the \ac{ASIC} area and energy results, and Section~\ref{sec:conclusions} concludes this work.

\section{Background} \label{sec:background}

\subsection{Posit Arithmetic}

Posit arithmetic is emerging as a promising general-purpose alternative to IEEE 754 floating-point numbers. Introduced in 2017 as a potential drop-in replacement for traditional floats~\cite{gustafson2017Beating}, posits have since attracted considerable interest across academia and industry. The Posit Number Standard~\cite{positworkinggroup2022Standard} specifies a posit configuration based on its total bit width, $n$, enabling the definition of posits of any size. Among these, the most commonly studied configurations in the literature are the byte-aligned posit8, posit16, posit32, and posit64 formats. The initial definition of posits~\cite{gustafson2017Beating} and earlier versions of the posit standard allowed for configuring the number of exponent bits. This was defined as \positformat{n}{es}, where the exponent size parameter \textit{es} enabled a trade-off between precision and dynamic range. Each increment in \textit{es} effectively doubled the dynamic range at the cost of one precision bit. However, in the most recent version of the standard~\cite{positworkinggroup2022Standard}, \textit{es} has been fixed to 2.

\begin{figure}[tbp]
    \centering
    \includegraphics[width=\columnwidth]{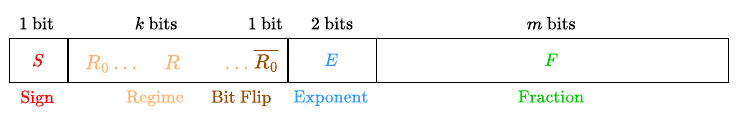}
    \caption{Posit format showing the sign, regime, exponent, and fraction fields.}
    \label{fig:posit_format}
\end{figure}

One of the key advantages of posit arithmetic is its simplicity, as it avoids the extensive range of special cases typically associated with floating-point arithmetic. Posits have only two special cases: the zero value, represented as $\texttt{0}\cdots\texttt{0}$, and the \ac{NaR}, represented as $\texttt{10}\cdots\texttt{0}$. All other bit patterns are used to encode values through the four distinct fields illustrated in Figure~\ref{fig:posit_format}.

The four bit fields of a posit are as follows:
\begin{itemize}
    \item The sign bit S encodes a value of $s=0$ if the number is positive or $s=1$ if the number is negative.
    \item The variable-length regime field R consists of a series of $k$ bits all equal to $R_0$ and is terminated by either $1-R_0$ or the end of the posit. The regime represents a long-range scaling factor $r$, which is calculated as:
    \begin{equation*}
        r = \left\{
    	\begin{array}{ll}
    		-k & \mbox{if } R_0 = 0 \\
    		k-1 & \mbox{if } R_0 = 1
    	\end{array}
    	\right.
    \end{equation*}
    \item The exponent field $E$ consists of at most $2$ bits and encodes an unbiased integer value $e$. Due to the variable-length nature of the regime field, one or both exponent bits may be located after the least significant bit of the posit, in which case these bits will be set to $0$.
    \item The fraction F is a variable-length field made up of the $m$ remaining bits. The value $f$ of the fraction is calculated by dividing the unsigned integer $F$ by $2^m$, ensuring that $0 \leq f < 1$.
\end{itemize}

The real value $p$ of a generic posit can be computed from its fields as follows:
\begin{equation} \label{eq:posit_value}
    p = ((1 - 3s) + f)\times 2^{(1-2s)\times(4r + e + s)}.
\end{equation}

This formula represents the most efficient method for decoding posits, as demonstrated by~\cite{murillo2022Comparing,uguen2019Evaluating}. The most notable differences in this value representation between posit arithmetic and the IEEE 754 floating-point standard are the existence of the variable-length regime, the use of an unbiased exponent, and the value of the hidden bits~\cite{murillo2022Comparing}.

\begin{figure}[tbp]
    \centering
    \includegraphics[width=\columnwidth]{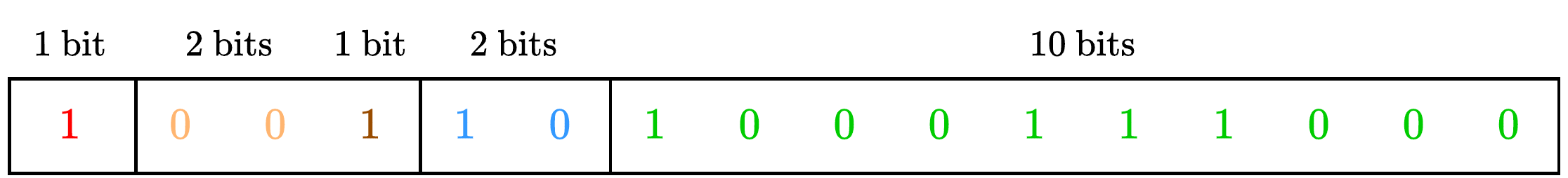}
    \caption{Decoding example of a posit16.}
    \label{fig:posit_format_example}
\end{figure}

As an example, let \texttt{1001101000111000} be the binary encoding of a Posit16 (Figure~\ref{fig:posit_format_example}). The first bit $s=\texttt{1}$ indicates a negative number. The regime field \texttt{001} gives $k=2$ and therefore $r=-2$. The next two bits \texttt{10} represent the exponent $e=2$. Finally, the remaining $m=10$ bits, \texttt{1000111000}, encode a fraction value of $f=568/2^{10}=0.5546875$. Hence, from Equation~(\ref{eq:posit_value}) we conclude that $\texttt{1001101000111000}\equiv (-2 + 0.5546875)\times 2^{-(4\cdot(-2)+2+1)} = -46.25$.

The variable-length regime field acts as a long-range dynamic exponent, as seen in Equation~(\ref{eq:posit_value}), where it is multiplied by four or, equivalently, shifted left by the two exponent bits. Since the regime and the fraction are dynamic fields, they allow for more flexibility in the trade-off between accuracy and dynamic range that a posit can achieve. If the regime field occupies more bits, it represents larger numbers at the cost of lower accuracy. On the other hand, when the regime field consists of fewer bits, posits have higher accuracy in the neighborhoods of $\pm 1$.

In posit arithmetic, \ac{NaR} has a unique representation that maps to the most negative 2's complement signed integer. Consequently, if used in comparison operations, it results in less than all other posits and is equal to itself. Moreover, the rest of the posit values follow the same ordering as their corresponding bit representations. These characteristics allow posit numbers to be compared as if they were 2's complement signed integers, eliminating additional hardware for posit comparison operations.

Posit arithmetic incorporates fused operations using the quire, a specialized fixed-point register with a width of $16n$ bits in 2's complement format. This accumulation register enables the execution of up to $2^{31} - 1$ \ac{MAC} operations without requiring intermediate rounding, thereby preserving numerical accuracy. Such fused operations are particularly advantageous in computations involving dot products, matrix multiplications, and other complex algorithms where precision is critical. However, the use of the quire imposes significant costs in terms of hardware area and power consumption~\cite{mallasen2022PERCIVAL}.

\begin{figure}[tbp]
    \centering
    \includegraphics[width=\columnwidth]{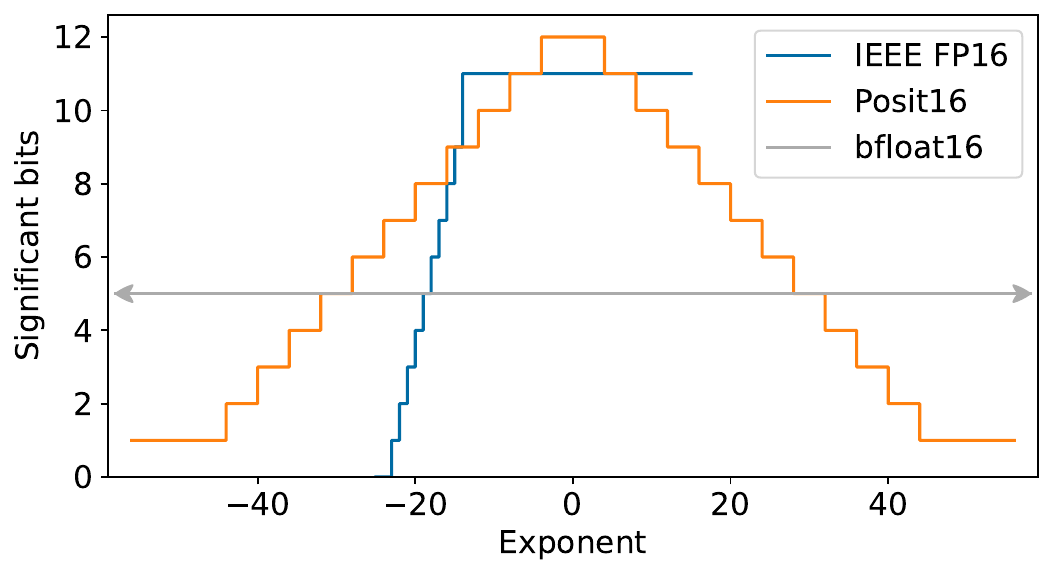}
    \caption{Accuracy and dynamic range of 16-bit arithmetic formats. The dynamic range of bfloat16 exceeds the limits of the plot.}
    \label{fig:16-bit_comparison}
\end{figure}

Of particular interest are low-precision posits, which present notable accuracy and dynamic range improvements over same-sized floats. This can be seen for 16-bit arithmetics in Figure~\ref{fig:16-bit_comparison}, where the maximum number of precision bits is 12 for posit16 and 11 for FP16. Regarding the dynamic range, the maximum reachable value of posit16 is $2^{56}\approx 7.21\times10^{16}$, while in the case of FP16, it is only $(2 - 2^{-10}) \times 2^{15} = 65520$. The case of bfloat16 is particular because it provides a significantly larger dynamic range (its maximum representable value is $\approx 3.4\times10^{38}$) but at the cost of having only 5 precision bits. As will be seen throughout this paper, the additional accuracy and dynamic range ratio provided by 8- to 16-bit posits enable the successful execution of algorithms using narrower posit configurations. This capability helps mitigate limitations related to memory bandwidth, offering a more efficient use of resources while maintaining computational precision.

\subsection{X-HEEP}

The eXtendible Heterogeneous Energy-Efficient Platform (X\nobreakdash-HEEP)~\cite{machetti2024XHEEP} is an open-source\footnote{\url{https://github.com/esl-epfl/x-heep}}, configurable framework designed to generate RISC\nobreakdash-V \acp{SoC} architectures and streamline the integration of tightly-coupled low-power edge accelerators. It provides extensive customization options, allowing developers to select CPU types, bus topologies, memories, and peripherals to suit specific application requirements.

The X\nobreakdash-HEEP platform leverages a robust ecosystem of high-quality open-source \acp{IP}, ensuring both efficiency and reliability in development. Key components include RISC\nobreakdash-V CPUs from the OpenHW Group, common \acp{IP} from the PULP platform, and peripherals from OpenTitan. Additionally, it incorporates custom-designed \acp{IP} such as I2S, DMA, and a power manager. By reusing silicon-proven IPs from the open-source community, X\nobreakdash-HEEP significantly reduces both design complexity and verification overhead. FuseSoC~\cite{fusesoc} powers the platform's simulation and synthesis workflows, enabling compatibility with \ac{FPGA} and \ac{ASIC} targets, thereby offering a flexible and scalable solution for diverse hardware implementations.

X\nobreakdash-HEEP is designed for easy extensibility through the eXtendible Accelerator InterFace (XAIF)~\cite{machetti2024XHEEP} and the \ac{CV-X-IF}~\cite{openhwgroupCoreV}, facilitating the deployment of custom hardware designs. Accelerators connected via the XAIF can efficiently interact with internal bus connections, signal the host CPU upon operation completion through interrupt ports, and manage power states using dedicated control ports. Meanwhile, the \ac{CV-X-IF} enables seamless CPU extension by allowing custom instructions to be implemented in a tightly coupled coprocessor without altering the original codebase of the CPU. X\nobreakdash-HEEP natively integrates \ac{CV-X-IF} support with the cv32e40x and cv32e40px CPUs. The cv32e40px is a fork of the cv32e40p~\cite{cv32e40p} extended with the \ac{CV-X-IF}. This flexibility accommodates diverse requirements in terms of area, power, and performance, making X\nobreakdash-HEEP a versatile and efficient platform for edge computing applications.

\section{Related Work} \label{sec:related_work}

Two primary strategies exist for using posit arithmetic as a substitute for floating-point computations in applications. The first strategy involves maintaining the same bit-width as the floating-point representation. This approach can significantly improve the application's computational accuracy but comes at the cost of increased system area and power consumption~\cite{mallasen2022PERCIVAL,mallasen2024BigPERCIVAL}. For instance, the enhanced accuracy achieved with posits has been demonstrated in scientific computing using the NAS Parallel Benchmark suite~\cite{chien2020Posit} and in the implementation of the CORDIC method~\cite{lim2020Approximating}. This is the main approach followed by applications that cannot tolerate inaccuracies in the computations.

The second strategy, which we have followed in this paper, focuses on reducing the bit-width of the data, accepting a minor accuracy trade-off to enhance system performance and efficiency. This approach has been applied in various domains, including weather and climate modeling, as explored in~\cite{klower2020Number}. In addition, it has been investigated for the training and inference of \acp{NN} using posits in~\cite{murillo2020Deep, raposo2021Positnn}, although integers have also been used for this purpose in low-power scenarios~\cite{burrello2022Bioformers}. When accuracy requirements allow, as is the case of~\cite{burrello2022Bioformers}, using an integer or fixed-point representation is generally more energy-efficient than floating point or posit.

On the hardware side, several previous proposals have integrated posit arithmetic capabilities into RISC\nobreakdash-V cores to varying extents. PERC~\cite{arunkumar2020PERC} and PERI~\cite{tiwari2021PERI} included \acp{PAU} into the Rocket core and the SHAKTI C-class core, respectively. However, these implementations were constrained by the RISC\nobreakdash-V F and D extensions, which prevented the inclusion of quire support. CLARINET~\cite{sharma2023CLARINET} added fused \ac{MAC}, and fused divide and accumulate operators using the quire to an RV32IMAFC RISC\nobreakdash-V core. PERCIVAL~\cite{mallasen2022PERCIVAL} provided complete support for posit32 with quire to the CVA6 core in parallel with pre-existing floating point capabilities thanks to the Xposit custom RISC\nobreakdash-V extension. This design was subsequently extended to posit64 in~\cite{mallasen2024BigPERCIVAL}. In a different approach, the authors of~\cite{cococcioni2022Lightweight} proposed performing the computations with an IEEE 754 \ac{FPU} but storing the values in memory using a lower bit-width posit format. These proposals had to modify the internal datapath of the base RISC\nobreakdash-V CPUs, in contrast to our proposed decoupled posit coprocessor described in Section~\ref{sec:PHEE}.

\section{Biomedical ML Applications} \label{sec:biomed_apps}

Biomedical \ac{ML}-based applications often demand a balance between computational efficiency and accuracy, especially in edge \ac{AI} wearable devices. This section examines the use of posits and compares them to other floating point representations, simulating the arithmetic formats using the Universal Numbers library~\cite{omtzigt2020Universal}, in two key applications: cough detection and R peak detection in \ac{ECG} signals. These case studies demonstrate how posit arithmetic can maintain high accuracy with reduced bit widths in error-tolerant scenarios, spanning both supervised and unsupervised learning contexts, offering a promising approach for enhancing the performance and efficiency of algorithms in resource-constrained biomedical devices.

\subsection{Cough Detection} \label{sec:cough_detection}

The first application we evaluated is a cough detection algorithm based on the one presented in~\cite{orlandic2023Multimodal}, which was later adapted to C for embedded platforms in~\cite{samakovlis2024BiomedBench}. This algorithm seeks to provide continuous monitoring of chronic cough patients in a privacy-preserving edge AI wearable device. The input biosignals are acquired using a 9-axis \ac{IMU} sampled at 100 Hz and encoded with 16 bits, and two microphones, each sampled at 16 kHz with 24-bit PCM encoding.

Each time window spans 300ms and is processed to decide if there has been a cough. The algorithm extracts several time-domain and frequency-domain features from both the \ac{IMU} and audio signals. The time domain features derived from the \ac{IMU} include the zero-crossing rate, kurtosis, and the root mean square. For the audio signals, the \ac{FFT} is used to obtain spectral statistics, power spectral density, and \acp{MFCC}. These features are then forwarded to a pre-trained random forest classifier, which calculates the probability of a cough event occurring within the time window.

The cough detection algorithm processes 200 random time windows that contain an equal amount of coughs, laughs, deep breaths, and throat clears from each of the 15 patients of the dataset from~\cite{orlandic2023Multimodal}. These results are then compared with the labels available in the dataset to measure the accuracy of the algorithm.

In this setup, we compared the impact of different arithmetic representations on the final accuracy of the application. The original algorithm employed 32-bit IEEE 754 floating-point numbers, but since the input signals are encoded with 16 or 24 bits, we also studied the use of half-precision floating-point numbers, posit16 with 2 and 3 exponent bits, posit24, posit32, and bfloat16.

\begin{figure}[tbp]
    \centering
    \includegraphics[width=\columnwidth]{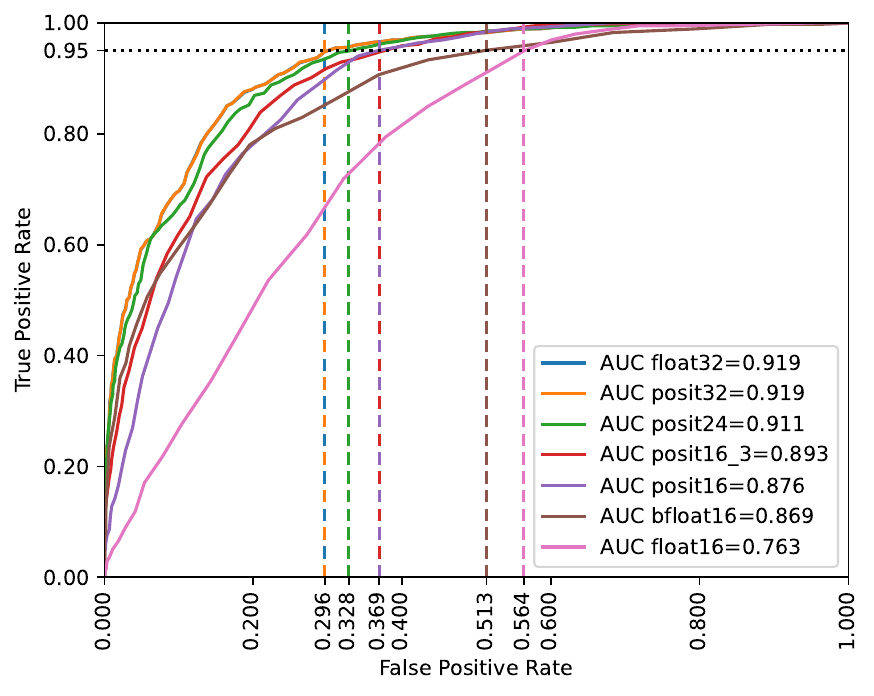}
    \caption{Cough detection accuracy results: \ac{ROC} curve for each arithmetic together with its \ac{AUC} and the \ac{FPR} for a \ac{TPR} of 0.95.}
    \label{fig:cough_roc_curve}
\end{figure}

For the comparison, following the methodology in~\cite{orlandic2023Multimodal}, we selected the \ac{ROC} curve and the \ac{AUC} metrics. The \ac{ROC} curve plots the \ac{TPR} against the \ac{FPR} at various threshold settings to evaluate the performance of a binary classification model. The \ac{AUC} summarizes this performance in a single scalar value ranging from 0 to 1. Figure~\ref{fig:cough_roc_curve} presents the \ac{ROC} curve along with the \ac{AUC} for each of the evaluated arithmetics. Additionally, it highlights the \ac{FPR} corresponding to a \ac{TPR} of 0.95. This metric would allow a physician to estimate the expected number of false positives when aiming to detect 95\% of a patient's coughs. A higher \ac{AUC} and a lower \ac{FPR} indicate improved accuracy for the respective arithmetic approach.

FP32 and posit32 obtain identical results with an \ac{AUC} value of 0.919 and an \ac{FPR} of 0.296. Notably, posit24 trails only slightly behind with an \ac{AUC} of 0.911 and \ac{FPR} of 0.328. The main differences between the arithmetic lie in the 16-bit range. Half-precision IEEE 754 numbers obtain the worst values for both metrics, only reaching an \ac{AUC} of 0.763 and an \ac{FPR} of 0.564. The bfloat16 format has a competitive \ac{AUC} of 0.869. However, its \ac{FPR} of 0.513 is close to that of FP16 and is significantly higher than that of the other arithmetics. Both posit16 and \positformat{16}{3} obtain an \ac{FPR} of 0.369, which represents a 34.6\% improvement over FP16 and a 28\% improvement over bfloat16. Furthermore, the non-standard \positformat{16}{3} has an \ac{AUC} of 0.893, while posit16 reaches 0.876. These values, contrary to FP16 or bfloat16, which are the go-to alternatives to FP32 in \ac{ML} applications, are relatively close to those of the 32-bit arithmetics.

This indicates that the cough detection algorithm could benefit from using 16-bit posits as a trade-off between high accuracy and high energy efficiency. In particular, when looking at the memory footprint, this application compiled with g++ and -O2 requires 629KB of memory when using FP32 and 447KB when using posit16. These values encompass the whole application requirements, including buffers, data, parameters, and code, resulting in a 29\% reduction in memory footprint.

The algorithm used in this study was designed with the main focus on benchmarking the performance of low-power wearable devices~\cite{samakovlis2024BiomedBench} rather than optimizing accuracy to the extent achieved by the method in~\cite{orlandic2023Multimodal}. The implementation in~\cite{orlandic2023Multimodal} was developed in Python with double-precision operations and a rich feature set to achieve high accuracy (AUC 0.90--0.97). In contrast, our study employed a C-based version optimized for microcontroller-class devices~\cite{samakovlis2024BiomedBench}, which introduced approximations such as table-based trigonometric functions and reduced feature sets to meet strict memory and runtime constraints. These adaptations naturally result in lower absolute accuracy (AUC 0.92 with 32-bit numbers), but this is consistent with the simplified embedded pipeline. Importantly, the goal of this work is not to surpass the accuracy of~\cite{orlandic2023Multimodal}, but to compare the relative performance of different arithmetic representations in a resource-constrained setting. This comparative analysis is independent of the absolute accuracy of the underlying biomedical application, ensuring the validity of our findings.

\subsection{R Peak Detection}

The BayeSlope algorithm was proposed in~\cite{degiovanni2023Adaptive} as a robust method for detecting R peaks, the basis for \ac{ECG} analysis, in edge \ac{AI} wearable sensors. BayeSlope was specially designed to be used under complex conditions, such as when performing intense physical exercise. It implements peak normalization through a generalized logistic function and a Bayesian filter to estimate the position of the following R peak. Then, it uses k-means clustering to divide the \ac{ECG} samples into a baseline centroid and a centroid representing the R peaks.

The authors of the BayeSlope method mention that they tried to replace the 32-bit floating-point implementation with a 32-bit fixed-point one. However, the fixed-point representation could not span the dynamic range required by the algorithm's clustering step, so they resorted to using FP32. Here, we will show that other floating-point representations with as few as 10 or even 8 bits can maintain the algorithm's high accuracy.

The dataset used to test BayeSlope comprises 20 subjects who performed an incremental test to exhaustion on a cycle ergometer~\cite{degiovanni2021ECG}. The algorithm analyzes windows of 1.75 seconds, with each of the five segments per subject in the dataset lasting approximately 25 seconds. The final accuracy tolerance of the R peak detection is set to the standard 150ms.

We ran the BayeSlope algorithm on the whole dataset with multiple arithmetic formats to assess their impact. The default IEEE 754 FP32 used by the authors of BayeSlope is compared to 32-bit posits, 16-bit numbers, and other low-bit-width formats, simulated using the Universal Numbers library. For the 16-bit case, we selected FP16, bfloat16, and posit16. As narrow arithmetic, we compared the non-standard 8-bit floating point numbers with 4 or 5 exponent bits---FP8E4M3 and FP8E5M2, respectively---, which have been proposed by industry for deep learning \acp{NN}~\cite{micikevicius2022FP8}, to posit12, posit10, and posit8.

\begin{figure}[tbp]
    \centering
    \includegraphics[width=\columnwidth]{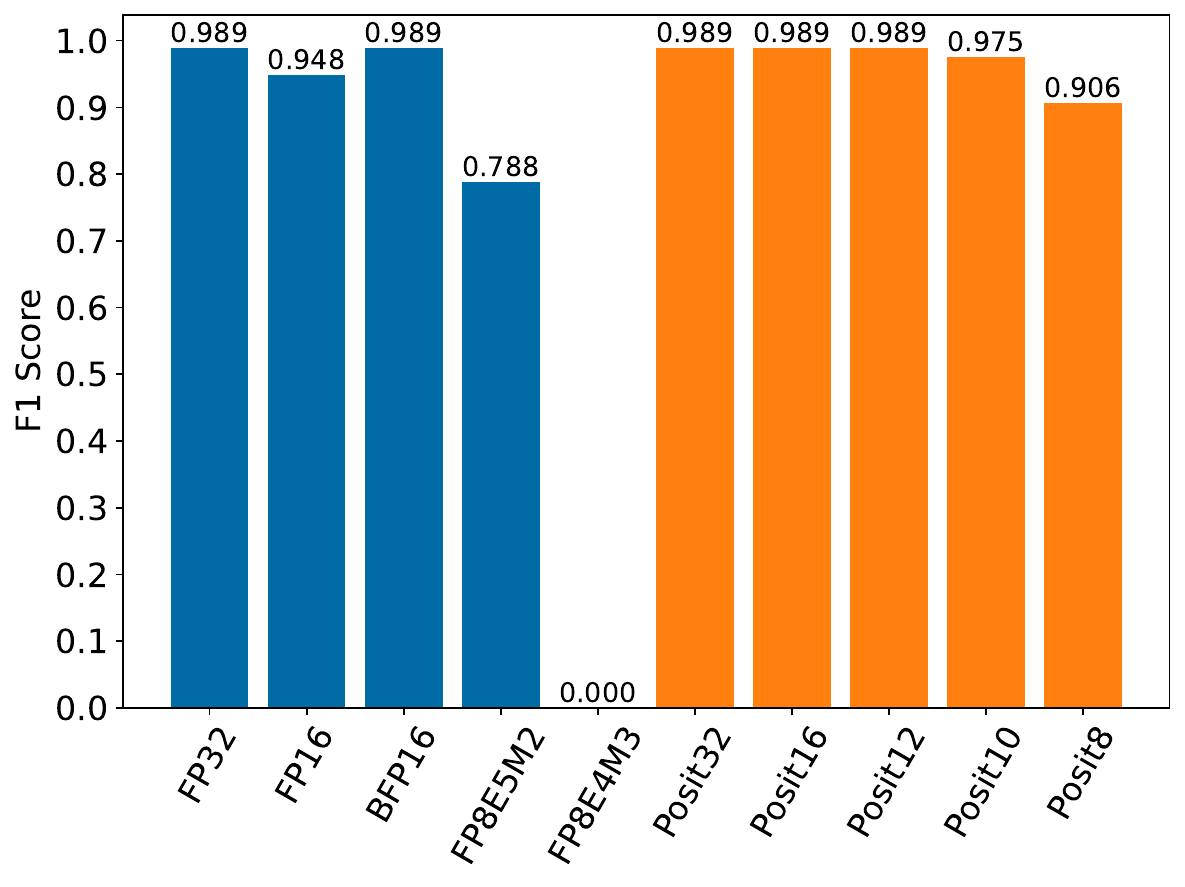}
    \caption{R peak detection accuracy results: F1 score for BayeSlope executed with different arithmetic formats.}
    \label{fig:bayeslope_f1_score}
\end{figure}

Figure~\ref{fig:bayeslope_f1_score} presents the F1 scores for different executions of BayeSlope using these arithmetic formats. As expected, both 32-bit formats achieve the highest F1 score of 0.989. In the 16-bit case, posit16 and bfloat16 maintain an identical F1 score, while FP16 performs poorly with a score of 0.948. This outcome suggests that the dynamic range of FP16 is insufficient in certain scenarios, underscoring the importance of the dynamic range over the arithmetic accuracy in BayeSlope.

\begin{figure}[tbp]
    \centering
    \includegraphics[width=\columnwidth]{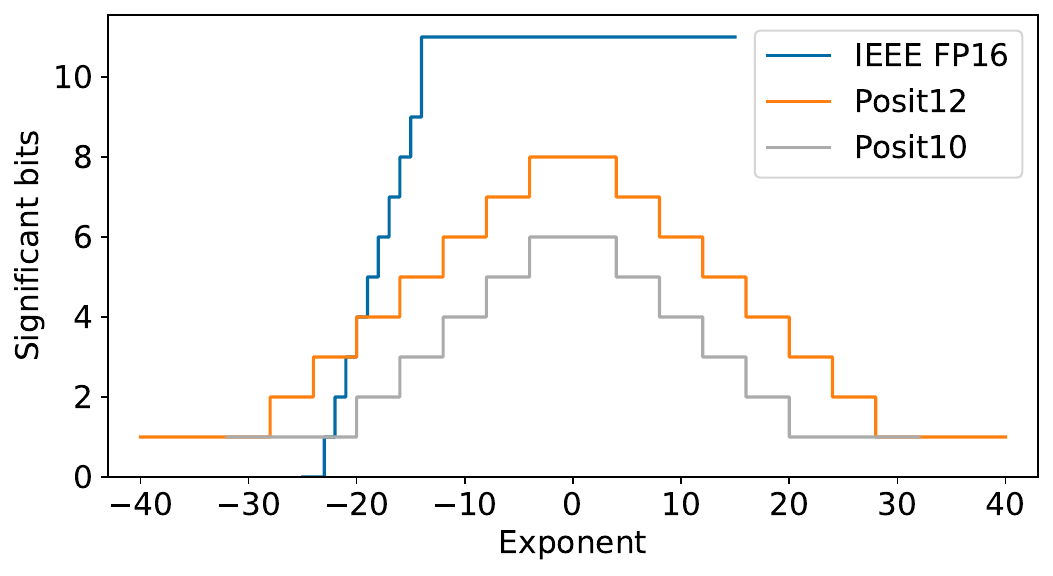}
    \caption{Accuracy and dynamic range of FP16, posit12, and posit10. The dynamic range of the posit formats is larger than that of FP16, although they have fewer significant bits.}
    \label{fig:12-bit_comparison}
\end{figure}

The performance of posit12 and posit10 further supports this observation. Despite their lower accuracy, these formats achieve higher F1 scores than FP16 due to their broader dynamic range (Figure~\ref{fig:12-bit_comparison}). Notably, posit12 matches the performance of the 32-bit formats, while posit10 falls only slightly short with an F1 score of 0.975.

Finally, 8-bit floating-point formats yield suboptimal results. FP8E4M3 lacks sufficient dynamic range to execute the algorithm successfully, and FP8E5M2 achieves a modest F1 score of 0.788. However, posit8 demonstrates competitive performance, achieving an F1 score of 0.906.

\section{PHEE} \label{sec:PHEE}

\begin{figure}[tbp]
    \centering
    \includegraphics[width=\columnwidth]{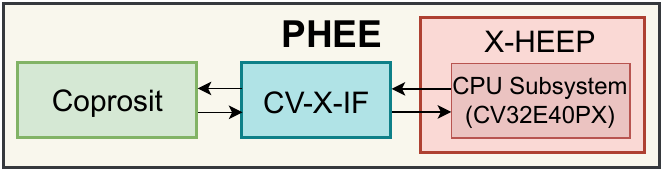}
    \caption{Block diagram of PHEE. Coprosit extends X-HEEP's CPU through the interface.}
    \label{fig:PHEE}
\end{figure}

To extend our evaluation from simulations to real-world hardware measurements, this section elaborates on the architectural design of PHEE\footnote{https://github.com/esl-epfl/PHEE}. This includes the design of Coprosit\footnote{https://github.com/esl-epfl/Coprosit}, our posit coprocessor specifically designed to integrate seamlessly with the \acf{CV-X-IF}. PHEE leverages the modularity and extensibility offered by the X-HEEP framework by embedding Coprosit into the cv32e40px RISC\nobreakdash-V core via the \ac{CV-X-IF} (Figure~\ref{fig:PHEE}). The selection of the cv32e40px core was motivated by two main factors: (1) It is well-suited for low-power edge AI applications, as it is a lightweight, energy-efficient RISC\nobreakdash-V core designed for microcontroller-class systems, making it representative of the class of devices targeted in this work. And (2) it provides native support for all the interfaces of the \ac{CV-X-IF} required by Coprosit and the FPU\_ss. The X-HEEP platform serves as the foundational hardware infrastructure, enabling the creation of a comprehensive \ac{SoC} that is synthesizable for both \acp{FPGA} and \acp{ASIC}.

As discussed in Section~\ref{sec:related_work}, PHEE adopts a novel architectural approach by decoupling all posit-related operations from the CPU's primary datapath. This design choice significantly improves modularity, enabling Coprosit to be easily reused and integrated into other systems that conform to the \ac{CV-X-IF} with minimal engineering overhead. Moreover, the separation of posit-related logic from the host CPU simplifies system maintainability. This decoupled architecture mitigates the need for frequent updates to accommodate internal changes within the CPU, thereby reducing the engineering effort required to sustain compatibility over time.

\subsection{Coprosit}

\begin{figure}[tbp]
    \centering
    \includegraphics[width=\columnwidth]{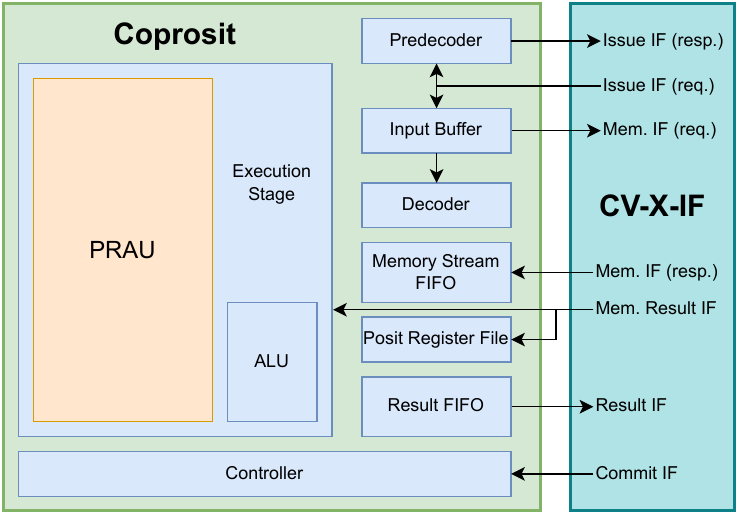}
    \caption{Block diagram of Coprosit. The execution stage contains the \ac{PRAU}, a small posit \ac{ALU}, and the blocks required to decouple all functionality from the CPU.}
    \label{fig:Coprosit}
\end{figure}

Coprosit is a configurable posit coprocessor designed to be fully compatible with the \ac{CV-X-IF}. Its architecture comprises the \ac{PRAU}, a lightweight \ac{ALU}, and the necessary components to control and communicate with the CPU via the interface (Figure~\ref{fig:Coprosit}).

\begin{figure}[tbp]
    \centering
    \includegraphics[width=\columnwidth]{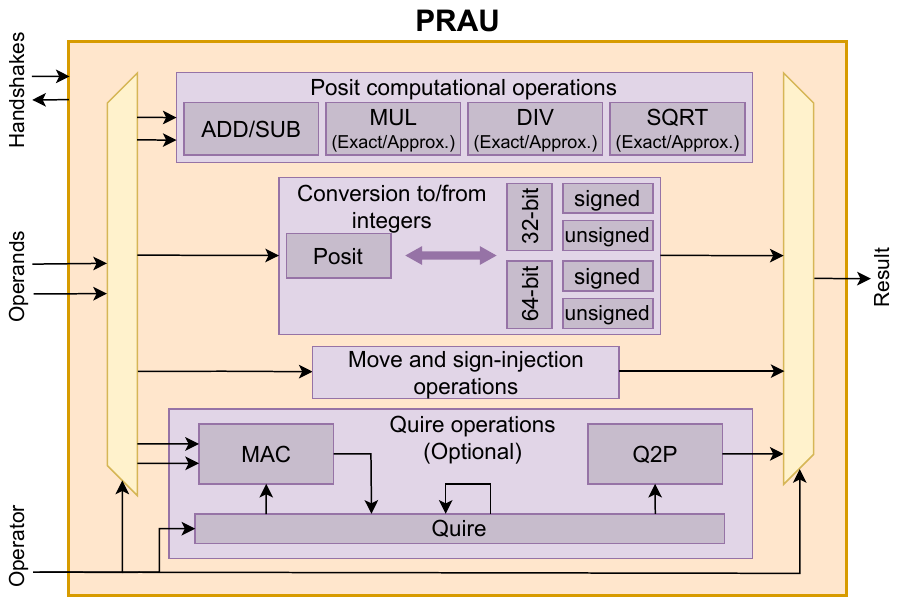}
    \caption{Block diagram of the \ac{PRAU}.}
    \label{fig:PRAU}
\end{figure}

The \ac{PRAU}\footnote{https://github.com/esl-epfl/PRAU}, inspired by the PAU's architecture described in~\cite{mallasen2024BigPERCIVAL}, serves as a standalone processing unit optimized for posit arithmetic (Figure~\ref{fig:PRAU}). This versatile unit operates using synchronous handshake protocols and supports either posit8, posit16, posit32, or posit64 formats, with optional quire functionality. These parameters are set statically when synthesizing the hardware. Its design aligns with the functional scope of the F and D standard RISC\nobreakdash-V extensions, which define the operations necessary for single- and double-precision IEEE 754 floating-point computations.

The \ac{PRAU} offers a comprehensive set of features, including:
\begin{itemize}
    \item Computational operations: Addition, subtraction, multiplication, division, and square root for posit numbers.
    \item Conversion operations: Transformations between posit representations and integers.
    \item Quire accumulator operations: clearing, negation, \ac{MAC}, and rounding to posit format.
    \item Register manipulation operations: Register moving and sign injection.
\end{itemize}

In addition to the \ac{PRAU}, the integrated ALU manages posit comparison operations. While such operations could theoretically be executed by the CPU's \ac{ALU} without the need for extra hardware~\cite{mallasen2022PERCIVAL}, doing so would undermine the architectural decoupling achieved through the dedicated interface.

The decoder and predecoder modules manage the offloading of instructions received via the \ac{CV-X-IF}, streamlining their processing. Additionally, Coprosit includes a 32-element posit register file and a dedicated controller, which ensure efficient and coordinated operation of the unit.

\section{ASIC Area and Energy Results} \label{sec:asic_results}

To evaluate the area and energy consumption of Coprosit in \ac{ASIC} synthesis, we targeted TSMC's 16~nm standard-cell library, operating at the typical corner of 0.8~V and 25ºC.  For comparison, we analyzed the results of Coprosit alongside those of FPU\_ss\footnote{\url{https://github.com/pulp-platform/fpu_ss}}, an architecture analogous to Coprosit but designed for IEEE 754 floating-point arithmetic, which utilizes the FPnew \ac{FPU}~\cite{mach2021FPnew}.

In alignment with the accuracy considerations outlined in Section~\ref{sec:biomed_apps}, Coprosit was statically configured to operate with posit16 arithmetic without quire functionality and with exact computational units, while FPU\_ss was configured for 32-bit floating point computation. This configuration enables a focused comparison of the use of each arithmetic in terms of area and energy efficiency within the scope of edge \ac{AI} wearables. Regarding other configuration parameters, the FPU\_ss has no ZFINX extension, so the coprocessors have their own floating-point registers. The input buffer depth was set to one, there was no out-of-order processing, and forwarding was enabled. These were required to ensure the correct execution of the operations in our tests. Finally, the FPU's features and implementation were set to their default values, which set combinational operations.

\subsection{Area Results}

The area results were obtained through \ac{ASIC} synthesis using Synopsys Design Compiler with a timing constraint of 2.35~ns. The critical path did not traverse the coprocessors but was in the CPU's communication through the system bus. Therefore, the clock period did not strain the implementation of either coprocessor. Table~\ref{tab:coprosit_fpuss_area} summarizes the area measurements for the individual modules within Coprosit and FPU\_ss. As expected, Coprosit exhibits a 38\% smaller area footprint compared to FPU\_ss. This reduction is primarily driven by the size differences in their core computational units: the \ac{PRAU} in Coprosit and the \ac{FPU} in FPU\_ss. Furthermore, the Coprosit register file contributes to the reduced area, as it is half the size of FPU\_ss’s register file due to the narrower data width.

\begin{table}[tbp]
    \centering
    \caption{Coprosit and FPU\_ss area results.}
    \label{tab:coprosit_fpuss_area}
    \begin{tabular}{@{}l
                        S[table-format=4.2]
                        S[table-format=4.2]
                    @{}}
    \toprule
        Area ($\mu m^2$)       & \text{Coprosit}   & \text{FPU\_ss}     \\ \midrule
        PRAU / FPU             & 2353.85           & 3726.26            \\
        Register File          & 878.79            & 1896.31            \\
        Controller             & 190.56            & 211.25             \\
        Input Buffer           & 178.33            & 231.41             \\
        Result FIFO            & 80.66             & {\qquad\text{---}} \\
        ALU                    & 79.11             & {\qquad\text{---}} \\
        Mem Stream FIFO        & 63.82             & 63.82              \\
        Decoder                & 31.52             & 25.87              \\
        Predecoder             & 9.07              & 11.20              \\ 
        CSR                    & {\qquad\text{---}} & 112.39            \\ 
        Compressed Predecoder  & {\qquad\text{---}} & 9.38              \\ \midrule
        Total                  & 4076.23           & 6565.43            \\ \bottomrule
    \end{tabular}%
\end{table}

The detailed comparisons in Table~\ref{tab:prau_fpu_area} further highlight this disparity. The 16-bit \ac{PRAU} is 37\% smaller than the 32-bit \ac{FPU}. In the \ac{PRAU}, the addition and multiplication operations are implemented as separate units that together occupy 576~$\mu m^2$. Conversely, the \ac{FPU} employs a fused multiply-accumulate (\ac{MAC}) unit for additions and multiplications, which requires 1800~$\mu m^2$, over three times the area of the corresponding units in the \ac{PRAU}.

\begin{table}[tbp]
    \centering
    \caption{PRAU and FPU area results in $\mu m^2$.}
    \label{tab:prau_fpu_area}
    \begin{tabular}{@{}l
                        S[table-format=4]
                        l
                        S[table-format=4]
                    @{}}
    \toprule
    PRAU        & 2354      & FPU & 3726   \\ \midrule
    Add         & 267       & \multirow{2}{*}{FMA}     & {\multirow{2}{*}{1800}} \\
    Mul         & 309       &                &            \\
    Sqrt        & 298       & \multirow{2}{*}{DivSqrt} & {\multirow{2}{*}{1078}}   \\
    Div         & 778       &                &            \\
    Conversions & 482       & Conversions    & 500        \\ \bottomrule
    \end{tabular}%
\end{table}

Table~\ref{tab:area_soa} summarizes the area results of the posit units from prior posit-enabled RISC\nobreakdash-V designs alongside our implementation. As shown, the PRAU in PHEE achieves a significantly smaller area footprint than prior designs, primarily due to its focus on posit16 and microcontroller-level edge-oriented constraints. For reference, in PHEE the CPU occupies 9750.43~$\mu m^2$. These differences reinforce the feasibility of implementing posit arithmetic in resource-constrained wearable systems.

\begin{table*}
    \centering
    \caption{Area comparison between posit units available in the literature.}
    \label{tab:area_soa}
    \begin{threeparttable}
    \begin{tabular}{@{}llllll@{}}
        \toprule
        Design   & Base core      & Posit format\tnote{a} & Quire & Technology & Area      \\ \midrule
        PERC~\cite{arunkumar2020PERC}     & Rocket Chip    & Posit32      & No    & FPGA (Spartan 7)       & 15949 LUT \\
        PERI~\cite{tiwari2021PERI}     & SHAKTI C-class & Posit32      & No    & TSMC 65 nm  & 74787.36~$\mu m^2$ \\
        CLARINET~\cite{sharma2023CLARINET} & Flute          & Posit32      & Yes   & TSMC 45 nm       & 69920.02~$\mu m^2$  \\
        Big-PERCIVAL~\cite{mallasen2024BigPERCIVAL} & CVA6           & Posit32      & No\tnote{b}   & TSMC 28 nm  & 18677.10~$\mu m^2$ \\
        PHEE (this work)  & cv32e40px      & Posit16      & No\tnote{b}   & TSMC 16 nm  & \;\;2432.96~$\mu m^2$\;\tnote{c} \\ \bottomrule
    \end{tabular}%
    \begin{tablenotes}
        \item [a] The designs allow for additional posit formats. Here, we show the smallest reported format.
        \item [b] The design also supports the quire, but the area results reported here are without.
        \item [c] Including the integer ALU for posit comparison operations.
    \end{tablenotes}
    \end{threeparttable}
\end{table*}

\subsection{Energy Results}

To ensure precise energy estimations, we obtained the switching activity of the signals in PHEE while running the main kernel of the cough detection application shown in Section~\ref{sec:cough_detection}: the \ac{FFT}. This kernel accounts for approximately 50\% of the algorithm’s runtime and is widely used as a preprocessing step of a range of \ac{ML}-based applications. This observation aligns with the findings in~\cite{samakovlis2024BiomedBench}, where the authors identify the computation of \acp{MFCC}---an operation involving iterative \ac{FFT} computations and transcendental functions---as the most computationally intensive kernel in the cough detection application. Using the Xposit compiler~\cite{mallasen2022PERCIVAL}, which natively supports posit arithmetic on PHEE, we were able to execute the \ac{FFT} kernel with posits and obtain post-synthesis energy measurements. These results serve as an additional metric to compare the use of these arithmetics in \ac{ML}-based biomedical applications.

The \ac{FFT} was implemented in C, producing three code versions: one for the posit implementation and two for the floating point implementation. In the posit version, arithmetic computations were handled using assembly instructions as required by the limitations of the Xposit compiler. For floating point, we implemented an assembly code identical to the posit one, with only negligible variations introduced later by the compiler in the optimization of loop orderings. This ensures consistency and fairness in the performance evaluation, and allows us to isolate the hardware contribution in our energy measurements. Furthermore, we implemented a floating-point version fully in C, allowing the compiler to optimize the code only for floating point. This approach allows us to highlight differences between the two hardware implementations of these arithmetics, while also showcasing the expected energy measurements in the playing field as it is today, where posits are not fully supported at the compilation level.

The three implementations were benchmarked using a 4096-element \ac{FFT}, which is comparable in size to the kernel used in the cough detection application. The posit-based implementation running on Coprosit required 1,495,623 clock cycles, while the floating-point implementation required 1,483,287 clock cycles in the assembly version, resulting in a mere 0.8\% difference, and 1,192,550 cycles in the non-assembly version, which results in a 20\% improvement in runtime. Since all computational operations in both implementations are combinational, the parity in execution time was expected when assembly was used in both cases.

To evaluate power consumption, we first obtained the dynamic switching activity of the post-synthesis simulation for the first two iterations of the \ac{FFT} kernel. These activity patterns were then used with Synopsys PrimePower to calculate power consumption values, as summarized in Table~\ref{tab:coprosit_fpuss_power} and Table~\ref{tab:prau_fpu_power}. In these tables and the following discussion, we only focused on the assembly versions to evaluate the hardware differences fairly.

Table~\ref{tab:coprosit_fpuss_power} details the power consumption of the main \ac{SoC} components, as well as individual modules in both Coprosit and FPU\_ss. The results show that Coprosit exhibits a total power consumption of 115~$\mu W$, which is approximately 28\% lower than FPU\_ss, whose total power consumption is 159~$\mu W$. The most significant contributor to this reduction is the lower power consumption of the \ac{PRAU} in Coprosit compared to the \ac{FPU} in FPU\_ss. Specifically, the \ac{PRAU} consumes 21.4~$\mu W$, which is 54\% less than the 46.5~$\mu W$ consumed by the \ac{FPU}. In our setup, where Coprosit is decoupled from the CPU, the PRAU would functionally also need the posit comparison instructions, for which we added the external lightweight ALU. Therefore, we could also account for the ALU's power consumption when comparing the \ac{PRAU} and the \ac{FPU}. In this case, the \ac{PRAU} + ALU requires 42.3\% less power than the FPU. Other modules, such as the input buffer and the register file, also demonstrate lower power consumption in Coprosit. However, Coprosit introduces additional modules, such as the Result FIFO and the ALU, which are absent in FPU\_ss. Despite these additions, Coprosit maintains the power advantage. At the \ac{SoC} level, the 512kB SRAM memory subsystem dominates overall power, and the CPU consumes around twice as much as the coprocessors. This emphasizes the necessity of reducing bit precision to reduce power consumption effectively. Lower bit-width formats not only reduce the coprocessor energy, but also shrink the memory footprint, as seen for the cough detection application in Section~\ref{sec:cough_detection}, and bandwidth requirements.

\begin{table}[tbp]
    \centering
    \caption{Detailed power values of Coprosit and FPU\_ss, as well as the power consumption of the CPU and memory subsystem of PHEE.}
    \label{tab:coprosit_fpuss_power}
    \begin{tabular}{@{}l
                        S[table-format=3.1]
                        S[table-format=3.1]
                    @{}}
    \toprule
        Power ($\mu W$)        & \text{Coprosit} & \text{FPU\_ss} \\ \midrule
        PRAU / FPU             & 21.4     & 46.5     \\
        Input Buffer           & 24.7     & 31.7     \\
        Regfile                & 19.1     & 29.9     \\
        Controller             & 16.3     & 16.6     \\
        Result FIFO            & 10.8     & \hfill\text{---}      \\
        Mem Stream FIFO        & 6.2      & 6.2     \\
        ALU                    & 5.4      & \hfill\text{---}      \\
        Decoder                & 1.1      & 1     \\
        Predecoder             & 0.3      & 0.4   \\
        CSR                    & \hfill\text{---} & 14.6     \\
        Compressed Predecoder  & \hfill\text{---} & 0.2     \\ \midrule
        Coprocessor total      & {\hfill115}      & {\hfill159} \\ \midrule
        cv32e40px CPU          & \multicolumn{2}{S[table-format=4.0]}{285} \\
        Memory\_ss             & \multicolumn{2}{S[table-format=4.0]}{1290}    \\ \bottomrule
    \end{tabular}%
\end{table}

Table~\ref{tab:prau_fpu_power} provides a more granular breakdown of the power consumption within the \ac{PRAU} and \ac{FPU}. The \ac{PRAU}'s addition and multiplication units consume 5.74~$\mu W$ and 1.32~$\mu W$, respectively. In contrast, the \ac{FPU}'s \ac{FMA} unit requires 36.1~$\mu W$, more than five times the combined consumption of the \ac{PRAU}'s addition and multiplication units. The power savings in the \ac{PRAU} also extend to the other operations. The division and square root units in the \ac{PRAU} together consume only 1.23~$\mu W$, compared to 5.42~$\mu W$ for the combined division and square root unit in the \ac{FPU}. Additionally, conversion operations are more efficient in the \ac{PRAU}, requiring just 0.13~$\mu W$, compared to 0.7~$\mu W$ in the \ac{FPU}. These large differences are partly due to the architectural differences between the \ac{PRAU} and the \ac{FPU}. In the \ac{PRAU}, the control signals are managed at the top level, whereas in the \ac{FPU}, this is taken care of within individual arithmetic units. This can be seen in Table~\ref{tab:prau_fpu_power} since the addition of the rows is far from the total in the case of the \ac{PRAU}, which indicates that the top-level module also accounts for a significant portion of the power value. Nonetheless, the overall 42.3\% decrease in power usage in the \ac{PRAU} + ALU highlights the advantages of using low bit-width data enabled by the characteristics of the posit representation.

\begin{table}[tbp]
    \centering
    \caption{PRAU and FPU detailed power values in $\mu W$.}
    \label{tab:prau_fpu_power}
    \begin{tabular}{@{}l
                        S[table-format=2.2]
                        l
                        S[table-format=2.2]
                    @{}}
    \toprule
    PRAU        & 21.4       & FPU & 46.5 \\ \midrule
    Add         & 5.74       & \multirow{2}{*}{FMA}      & {\multirow{2}{*}{36.1}\hfill} \\
    Mul         & 1.32       &                &            \\
    Sqrt        & 0.37       & \multirow{2}{*}{DivSqrt} & {\hfill\multirow{2}{*}{5.42}} \\
    Div         & 0.86       &                &            \\
    Conversions & 0.13       & Conversions    & 0.7 \\ \bottomrule
    \end{tabular}%
\end{table}

Combining these observations, we calculated the energy differences across implementations. When executing the non-assembly floating-point version, while there was a decrease in runtime, the power consumption of the FPU\_ss rose from 159~$\mu W$ to 179~$\mu W$ due to increased  switching activity. Translating these figures into energy consumption, Coprosit consumes 404.2~nJ, compared to 554.2~nJ for FPU\_ss with assembly and 501.6~nJ for FPU\_ss without assembly, assuming a 2.35~ns clock period. When isolating hardware differences in the assembly-based implementations, Coprosit achieves a 27.1\% energy reduction relative to FPU\_ss. In the current state of posit arithmetic, where compilation support is still limited, Coprosit still consumes 19.4\% less energy than the FPU\_ss in the non-assembly version. These findings highlight the energy benefits of leveraging low-precision arithmetic formats, reinforcing their potential for future energy-efficient edge \ac{AI} wearable systems.

\section{Conclusions} \label{sec:conclusions}

Energy consumption is a critical metric in wearable edge \ac{AI} biomedical devices, directly influencing their usability. This study has analyzed the impact of arithmetic representations on the \ac{ML}-based algorithms executed on these devices. Our results show that while formats like half-precision floats and bfloat16 lead to a significant accuracy drop, 16-bit posits can replace 32-bit IEEE 754 floating-point numbers in a cough detection application with minimal accuracy degradation. Extending this analysis to R peak detection in \acp{ECG}, we found that while floating-point representations require a minimum of 16 bits for satisfactory accuracy, posit arithmetic allows for reductions to 10 or even 8 bits, maintaining acceptable performance.

The hardware implementation of PHEE, integrating the Coprosit posit coprocessor via the \ac{CV-X-IF}, enabled area and post-synthesis energy measurements targeting 16~nm TSMC technology. Comparative analysis between the posit16 and single-precision IEEE 754 formats, guided by the precision requirements of the target applications, revealed that posit hardware achieves a 38\% reduction in area compared to a single-precision \ac{FPU} equivalent design. Furthermore, while both formats exhibit similar performance, the posit hardware consumes up to 42.3\% less power at the functional unit level. When taking into account the host CPU and the system memory, the impact of posit power is marginal, of less than 7\%. Yet, it can save up to 29\% of the memory footprint at the application level and between 19 and 27\% of energy at the coprocessor level compared to a floating-point implementation.

Prior research has demonstrated that, when using representations of the same size, posits incur a higher area and power cost than traditional floating-point arithmetic. Consequently, floating-point should be the preferred choice when accuracy is not a limiting factor in the target application. Furthermore, whenever fixed-point can meet the accuracy requirements of an application, it is generally the most energy-efficient option. However, this study shows that certain \ac{ML}-based biomedical applications are constrained by accuracy requirements, both using supervised and unsupervised learning. In such cases, a narrower posit representation offers a viable intermediate solution. It provides greater precision than a narrower floating-point format, which may be insufficient, while being more efficient than a larger, more costly floating-point alternative. These findings underscore the potential of low-precision posit arithmetic to significantly enhance the energy efficiency of edge \ac{AI} wearables, making it a promising approach for future designs.

\section*{Acknowledgments}

This work was supported in part by grants PID2021-123041OB-I00 and PID2021-126576NB-I00 funded by MCIN/AEI/ 10.13039/501100011033 and by “ERDF A way of making Europe”, and supported in part by the Swiss State Secretariat for Education, Research and Innovation (SERI) through the SwissChips research project. 

\bibliographystyle{IEEEtran}
\bibliography{references}

@inproceedings{orlandic2023Multimodal,
  title = {A {{Multimodal Dataset}} for {{Automatic Edge-AI Cough Detection}}},
  booktitle = {2023 45th {{Annual International Conference}} of the {{IEEE Engineering}} in {{Medicine}} \& {{Biology Society}} ({{EMBC}})},
  author = {Orlandic, Lara and Thevenot, J{\'e}r{\^o}me and Teijeiro, Tomas and Atienza, David},
  year = {2023},
  month = jul,
  pages = {1--7},
  publisher = {{IEEE}},
  address = {{Sydney, Australia}},
  doi = {10.1109/EMBC40787.2023.10340413},
  urldate = {2023-12-27},
  isbn = {9798350324471},
  langid = {english}
}

@misc{machetti2024XHEEP,
  title = {X-{{HEEP}}: {{An Open-Source}}, {{Configurable}} and {{Extendible RISC-V Microcontroller}} for the {{Exploration}} of {{Ultra-Low-Power Edge Accelerators}}},
  shorttitle = {X-{{HEEP}}},
  author = {Machetti, Simone and Schiavone, Pasquale Davide and M{\"u}ller, Thomas Christoph and {Pe{\'o}n-Quir{\'o}s}, Miguel and Atienza, David},
  year = {2024},
  month = jan,
  number = {arXiv:2401.05548},
  eprint = {2401.05548},
  primaryclass = {cs},
  publisher = {{arXiv}},
  urldate = {2024-01-16},
  archiveprefix = {arxiv},
  langid = {english}
}

@article{degiovanni2023Adaptive,
  title = {Adaptive {{R-Peak Detection}} on {{Wearable ECG Sensors}} for {{High-Intensity Exercise}}},
  author = {De Giovanni, Elisabetta and Teijeiro, Tomas and Millet, Gregoire P. and Atienza, David},
  year = {2023},
  month = mar,
  journal = {IEEE Transactions on Biomedical Engineering},
  volume = {70},
  number = {3},
  pages = {941--953},
  issn = {0018-9294, 1558-2531},
  doi = {10.1109/TBME.2022.3205304},
  urldate = {2024-02-14},
  langid = {english}
}

@article{samakovlis2024BiomedBench,
  title = {{{BiomedBench}}: {{A}} Benchmark Suite of {{TinyML}} Biomedical Applications for Low-Power Wearables},
  shorttitle = {{{BiomedBench}}},
  author = {Samakovlis, Dimitrios and Albini, Stefano and {\'A}lvarez, Rub{\'e}n Rodr{\'i}guez and Constantinescu, Denisa-Andreea and Schiavone, Pasquale Davide and {Pe{\'o}n-Quir{\'o}s}, Miguel and Atienza, David},
  year = {2024},
  journal = {IEEE Design \& Test},
  pages = {1--1},
  issn = {2168-2364},
  doi = {10.1109/MDAT.2024.3483034},
  urldate = {2025-01-21}
}

@article{omtzigt2020Universal,
  title = {Universal {{Numbers Library}}: Design and Implementation of a High-Performance Reproducible Number Systems Library},
  author = {Omtzigt, E. Theodore L. and Gottschling, Peter and Seligman, Mark and Zorn, William},
  year = {2020},
  journal = {arXiv:2012.11011},
  eprint = {2012.11011},
  eprinttype = {arxiv},
  archiveprefix = {arXiv}
}

@misc{positworkinggroup2022Standard,
  title = {Standard for {{Posit Arithmetic}} (2022)},
  shorttitle = {Standard for {{Posit Arithmetic}} (2022)},
  author = {{Posit Working Group}},
  year = {2022},
  month = feb,
  urldate = {2022-05-16},
  howpublished = {https://posithub.org/docs/posit\_standard-2.pdf}
}

@incollection{murillo2022Comparing,
  title = {Comparing {{Different Decodings}} for {{Posit Arithmetic}}},
  booktitle = {Next {{Generation Arithmetic}}},
  author = {Murillo, Raul and Mallas{\'e}n, David and Del Barrio, Alberto A. and Botella, Guillermo},
  editor = {Gustafson, John and Dimitrov, Vassil},
  year = {2022},
  volume = {13253},
  pages = {84--99},
  publisher = {Springer International Publishing},
  address = {Cham},
  doi = {10.1007/978-3-031-09779-9_6},
  urldate = {2022-07-20},
  isbn = {978-3-031-09778-2 978-3-031-09779-9},
  langid = {english}
}

@inproceedings{uguen2019Evaluating,
  title = {Evaluating the {{Hardware Cost}} of the {{Posit Number System}}},
  booktitle = {2019 29th {{International Conference}} on {{Field Programmable Logic}} and {{Applications}} ({{FPL}})},
  author = {Uguen, Yohann and Forget, Luc and {de Dinechin}, Florent},
  year = {2019},
  month = sep,
  pages = {106--113},
  publisher = {IEEE},
  address = {Barcelona, Spain},
  doi = {10.1109/FPL.2019.00026},
  urldate = {2021-10-07},
  isbn = {978-1-72814-884-7}
}

@article{gustafson2017Beating,
  title = {Beating Floating Point at Its Own Game: {{Posit}} Arithmetic},
  author = {Gustafson, John L. and Yonemoto, Isaac T.},
  year = {2017},
  month = apr,
  journal = {Supercomputing Frontiers and Innovations},
  volume = {4},
  number = {2},
  pages = {71--86},
  doi = {10.14529/jsfi170206}
}

@misc{openhwgroupCoreV,
  title = {Core-{{V eXtension}} Interface ({{CV-X-IF}}) v1.0.0 Documentation},
  author = {{OpenHW Group}},
  urldate = {2024-05-28},
  year = {2024},
  howpublished = {\url{https://docs.openhwgroup.org/projects/openhw-group-core-v-xif/en/v1.0.0/}}
}

@misc{fusesoc,
  title = {FuseSoC},
  author = {{Olof Kindgren}},
  urldate = {2025-01-27},
  year = {2025},
  howpublished = {\url{https://github.com/olofk/fusesoc}}
}

@article{cv32e40p,
  title={Near-threshold RISC-V core with DSP extensions for scalable IoT endpoint devices},
  author={Gautschi, Michael and Schiavone, Pasquale Davide and Traber, Andreas and Loi, Igor and Pullini, Antonio and Rossi, Davide and Flamand, Eric and G{\"u}rkaynak, Frank K and Benini, Luca},
  journal={IEEE transactions on very large scale integration (VLSI) systems},
  volume={25},
  number={10},
  pages={2700--2713},
  year={2017},
  publisher={IEEE}
}

@article{klower2020Number,
  title = {Number {{Formats}}, {{Error Mitigation}}, and {{Scope}} for 16-{{Bit Arithmetics}} in {{Weather}} and {{Climate Modeling Analyzed With}} a {{Shallow Water Model}}},
  author = {Kl{\"o}wer, M. and D{\"u}ben, P. D. and Palmer, T. N.},
  year = {2020},
  journal = {Journal of Advances in Modeling Earth Systems},
  volume = {12},
  number = {10},
  pages = {e2020MS002246},
  issn = {1942-2466},
  doi = {10.1029/2020MS002246},
  urldate = {2023-02-14},
  langid = {english}
}

@article{mallasen2022PERCIVAL,
  title = {{{PERCIVAL}}: {{Open-Source Posit RISC-V Core With Quire Capability}}},
  shorttitle = {{{PERCIVAL}}},
  author = {Mallas{\'e}n, David and Murillo, Raul and Barrio, Alberto A. Del and Botella, Guillermo and Pi{\~n}uel, Luis and {Prieto-Matias}, Manuel},
  year = {2022},
  month = jul,
  journal = {IEEE Transactions on Emerging Topics in Computing},
  volume = {10},
  number = {3},
  pages = {1241--1252},
  issn = {2168-6750, 2376-4562},
  doi = {10.1109/TETC.2022.3187199},
  urldate = {2024-03-20},
  langid = {english}
}

@article{mallasen2024BigPERCIVAL,
  title = {Big-{{PERCIVAL}}: {{Exploring}} the {{Native Use}} of 64-{{Bit Posit Arithmetic}} in {{Scientific Computing}}},
  shorttitle = {Big-{{PERCIVAL}}},
  author = {Mallas{\'e}n, David and Del Barrio, Alberto A. and {Prieto-Matias}, Manuel},
  year = {2024},
  month = jun,
  journal = {IEEE Transactions on Computers},
  volume = {73},
  number = {6},
  pages = {1472--1485},
  issn = {0018-9340, 1557-9956, 2326-3814},
  doi = {10.1109/TC.2024.3377890},
  urldate = {2024-05-22},
  copyright = {https://creativecommons.org/licenses/by-nc-nd/4.0/},
  langid = {english}
}

@article{mach2021FPnew,
  title = {{{FPnew}}: {{An Open-Source Multiformat Floating-Point Unit Architecture}} for {{Energy-Proportional Transprecision Computing}}},
  shorttitle = {{{FPnew}}},
  author = {Mach, Stefan and Schuiki, Fabian and Zaruba, Florian and Benini, Luca},
  year = {2021},
  month = apr,
  journal = {IEEE Transactions on Very Large Scale Integration (VLSI) Systems},
  volume = {29},
  number = {4},
  pages = {774--787},
  issn = {1557-9999},
  doi = {10.1109/TVLSI.2020.3044752}
}

@misc{degiovanni2021ECG,
  title = {{{ECG}} in High Intensity Exercise Dataset},
  author = {De Giovanni, Elisabetta and Teijeiro, Tomas and Meier, David and Millet, Gr{\'e}goire and Atienza, David},
  year = {2021},
  month = nov,
  publisher = {Zenodo},
  doi = {10.5281/zenodo.5727800}
}

@article{chong2019Energy,
  title = {Energy {{Harvesting For Wearable Devices}}: {{A Review}}},
  shorttitle = {Energy {{Harvesting For Wearable Devices}}},
  author = {Chong, Yung-Wey and Ismail, Widad and Ko, Kwangman and Lee, Chen-Yi},
  year = {2019},
  month = oct,
  journal = {IEEE Sensors Journal},
  volume = {19},
  number = {20},
  pages = {9047--9062},
  issn = {1558-1748},
  doi = {10.1109/JSEN.2019.2925638},
  urldate = {2025-01-09}
}

@inproceedings{ferretti2022INCLASS,
  title = {{{INCLASS}}: {{Incremental Classification Strategy}} for {{Self-Aware Epileptic Seizure Detection}}},
  shorttitle = {{{INCLASS}}},
  booktitle = {2022 {{Design}}, {{Automation}} \& {{Test}} in {{Europe Conference}} \& {{Exhibition}} ({{DATE}})},
  author = {Ferretti, Lorenzo and Ansaloni, Giovanni and Marquis, Renaud and Teijeiro, Tomas and Ryvlin, Philippe and Atienza, David and Pozzi, Laura},
  year = {2022},
  month = mar,
  pages = {1449--1454},
  publisher = {IEEE},
  address = {Antwerp, Belgium},
  doi = {10.23919/DATE54114.2022.9774713},
  urldate = {2023-12-27},
  isbn = {978-3-9819263-6-1},
  langid = {english}
}

@article{wang2024ACE,
  title = {{{ACE}}: {{Automated}} Optimization towards Iterative {{Classification}} in {{Edge}} Health Monitors},
  shorttitle = {{{ACE}}},
  author = {Wang, Yuxuan and Orlandic, Lara and Machetti, Simone and Ansaloni, Giovanni and Atienza, David},
  year = {2024},
  journal = {IEEE Transactions on Biomedical Circuits and Systems},
  pages = {1--11},
  issn = {1940-9990},
  doi = {10.1109/TBCAS.2024.3468160},
  urldate = {2025-01-09}
}

@article{momeni2022CAFS,
  title = {{{CAFS}}: {{Cost-Aware Features Selection Method}} for {{Multimodal Stress Monitoring}} on {{Wearable Devices}}},
  shorttitle = {{{CAFS}}},
  author = {Momeni, Niloofar and Vald{\'e}s, Adriana Arza and Rodrigues, Jo{\~a}o and Sandi, Carmen and Atienza, David},
  year = {2022},
  month = mar,
  journal = {IEEE Transactions on Biomedical Engineering},
  volume = {69},
  number = {3},
  pages = {1072--1084},
  issn = {1558-2531},
  doi = {10.1109/TBME.2021.3113593},
  urldate = {2025-01-09}
}

@incollection{gunaratne2023Evaluation,
  title = {Evaluation of the {{Use}} of {{Low Precision Floating-Point Arithmetic}} for {{Applications}} in {{Radio Astronomy}}},
  booktitle = {Next {{Generation Arithmetic}}},
  author = {Gunaratne, Thushara Kanchana},
  editor = {Gustafson, John and Leong, Siew Hoon and Michalewicz, Marek},
  year = {2023},
  volume = {13851},
  pages = {155--170},
  publisher = {Springer Nature Switzerland},
  address = {Cham},
  doi = {10.1007/978-3-031-32180-1_10},
  urldate = {2023-05-15},
  isbn = {978-3-031-32179-5 978-3-031-32180-1},
  langid = {english}
}

@incollection{chien2020Posit,
  title = {Posit {{NPB}}: {{Assessing}} the {{Precision Improvement}} in {{HPC Scientific Applications}}},
  shorttitle = {Posit {{NPB}}},
  booktitle = {Parallel {{Processing}} and {{Applied Mathematics}}},
  author = {Chien, Steven W. D. and Peng, Ivy B. and Markidis, Stefano},
  editor = {Wyrzykowski, Roman and Deelman, Ewa and Dongarra, Jack and Karczewski, Konrad},
  year = {2020},
  volume = {12043},
  pages = {301--310},
  publisher = {Springer International Publishing},
  address = {Cham},
  doi = {10.1007/978-3-030-43229-4_26},
  urldate = {2021-12-02},
  isbn = {978-3-030-43228-7 978-3-030-43229-4},
  langid = {english}
}

@article{murillo2020Deep,
  title = {Deep {{PeNSieve}}: {{A}} Deep Learning Framework Based on the Posit Number System},
  shorttitle = {Deep {{PeNSieve}}},
  author = {Murillo, Raul and Del Barrio, Alberto A. and Botella, Guillermo},
  year = {2020},
  month = jul,
  journal = {Digital Signal Processing},
  volume = {102},
  pages = {102762},
  issn = {10512004},
  doi = {10.1016/j.dsp.2020.102762},
  urldate = {2021-10-30},
  langid = {english}
}

@inproceedings{raposo2021Positnn,
  title = {Positnn: {{Training Deep Neural Networks}} with {{Mixed Low-Precision Posit}}},
  shorttitle = {Positnn},
  booktitle = {{{ICASSP}} 2021 - 2021 {{IEEE International Conference}} on {{Acoustics}}, {{Speech}} and {{Signal Processing}} ({{ICASSP}})},
  author = {Raposo, Gon{\c c}alo and Tom{\'a}s, Pedro and Roma, Nuno},
  year = {2021},
  month = jun,
  pages = {7908--7912},
  issn = {2379-190X},
  doi = {10.1109/ICASSP39728.2021.9413919}
}

@inproceedings{horowitz201411,
  title = {1.1 {{Computing}}'s Energy Problem (and What We Can Do about It)},
  booktitle = {2014 {{IEEE International Solid-State Circuits Conference Digest}} of {{Technical Papers}} ({{ISSCC}})},
  author = {Horowitz, Mark},
  year = {2014},
  month = feb,
  pages = {10--14},
  issn = {2376-8606},
  doi = {10.1109/ISSCC.2014.6757323},
  urldate = {2025-01-16}
}

@inproceedings{arunkumar2020PERC,
  title = {{{PERC}}: {{Posit Enhanced Rocket Chip}}},
  booktitle = {4th {{Workshop}} on {{Computer Architecture Research}} with {{RISC-V}} ({{CARRV}}'20)},
  author = {Arunkumar, M. V. and Bhairathi, Sai Ganesh and Hayatnagarkar, Harshal G.},
  year = {2020},
  pages = {8},
  langid = {english}
}

@article{tiwari2021PERI,
  title = {{{PERI}}: {{A Configurable Posit Enabled RISC-V Core}}},
  shorttitle = {{{PERI}}},
  author = {Tiwari, Sugandha and Gala, Neel and Rebeiro, Chester and Kamakoti, V.},
  year = {2021},
  month = jun,
  journal = {ACM Transactions on Architecture and Code Optimization},
  volume = {18},
  number = {3},
  pages = {1--26},
  issn = {1544-3566, 1544-3973},
  doi = {10.1145/3446210},
  urldate = {2021-09-26},
  langid = {english}
}

@article{sharma2023CLARINET,
  title = {{{CLARINET}}: {{A}} Quire-Enabled {{RISC-V-based}} Framework for Posit Arithmetic Empiricism},
  shorttitle = {{{CLARINET}}},
  author = {Sharma, Niraj N. and Jain, Riya and Pokkuluri, Mohana Madhumita and Patkar, Sachin B. and Leupers, Rainer and Nikhil, Rishiyur S. and Merchant, Farhad},
  year = {2023},
  month = feb,
  journal = {Journal of Systems Architecture},
  volume = {135},
  pages = {102801},
  issn = {13837621},
  doi = {10.1016/j.sysarc.2022.102801},
  urldate = {2024-03-21},
  langid = {english}
}

@article{cococcioni2022Lightweight,
  title = {A {{Lightweight Posit Processing Unit}} for {{RISC-V Processors}} in {{Deep Neural Network Applications}}},
  author = {Cococcioni, Marco and Rossi, Federico and Ruffaldi, Emanuele and Saponara, Sergio},
  year = {2022},
  month = oct,
  journal = {IEEE Transactions on Emerging Topics in Computing},
  volume = {10},
  number = {4},
  pages = {1898--1908},
  issn = {2168-6750, 2376-4562},
  doi = {10.1109/TETC.2021.3120538},
  urldate = {2024-06-11},
  copyright = {https://ieeexplore.ieee.org/Xplorehelp/downloads/license-information/IEEE.html},
  langid = {english}
}

@inproceedings{lim2020Approximating,
  title = {Approximating Trigonometric Functions for Posits Using the {{CORDIC}} Method},
  booktitle = {Proceedings of the 17th {{ACM International Conference}} on {{Computing Frontiers}}},
  author = {Lim, Jay P. and Shachnai, Matan and Nagarakatte, Santosh},
  year = {2020},
  month = may,
  pages = {19--28},
  publisher = {ACM},
  address = {Catania Sicily Italy},
  doi = {10.1145/3387902.3392632},
  urldate = {2023-04-17},
  isbn = {978-1-4503-7956-4},
  langid = {english}
}

@article{wei2020Review,
  title = {A {{Review}} of {{Algorithm}} \& {{Hardware Design}} for {{AI-Based Biomedical Applications}}},
  author = {Wei, Ying and Zhou, Jun and Wang, Yin and Liu, Yinggang and Liu, Qingsong and Luo, Jiansheng and Wang, Chao and Ren, Fengbo and Huang, Li},
  year = {2020},
  month = apr,
  journal = {IEEE Transactions on Biomedical Circuits and Systems},
  volume = {14},
  number = {2},
  pages = {145--163},
  issn = {1940-9990},
  doi = {10.1109/TBCAS.2020.2974154},
  urldate = {2025-01-27}
}

@article{liu2023UltraLow,
  title = {An {{Ultra-Low Power Reconfigurable Biomedical AI Processor With Adaptive Learning}} for {{Versatile Wearable Intelligent Health Monitoring}}},
  author = {Liu, Jiahao and Fan, Jiajing and Zhong, Zirui and Qiu, Hui and Xiao, Jianbiao and Zhou, Yong and Zhu, Zhen and Dai, Guanghai and Wang, Ning and Liu, Qingsong and Xie, Yuxiang and Liu, Hongduo and Chang, Liang and Zhou, Jun},
  year = {2023},
  month = oct,
  journal = {IEEE Transactions on Biomedical Circuits and Systems},
  volume = {17},
  number = {5},
  pages = {952--967},
  issn = {1940-9990},
  doi = {10.1109/TBCAS.2023.3276782},
  urldate = {2025-01-27}
}

@inproceedings{burrello2022Bioformers,
  title = {Bioformers: {{Embedding Transformers}} for {{Ultra-Low Power sEMG-based Gesture Recognition}}},
  shorttitle = {Bioformers},
  booktitle = {2022 {{Design}}, {{Automation}} \& {{Test}} in {{Europe Conference}} \& {{Exhibition}} ({{DATE}})},
  author = {Burrello, Alessio and Morghet, Francesco Bianco and Scherer, Moritz and Benatti, Simone and Benini, Luca and Macii, Enrico and Poncino, Massimo and Pagliari, Daniele Jahier},
  year = {2022},
  month = mar,
  pages = {1443--1448},
  issn = {1558-1101},
  doi = {10.23919/DATE54114.2022.9774639},
  urldate = {2025-01-27}
}

@article{dekimpe2022ECG,
  title = {{{ECG Arrhythmia Classification}} on an {{Ultra-Low-Power Microcontroller}}},
  author = {Dekimpe, R{\'e}mi and Bol, David},
  year = {2022},
  month = jun,
  journal = {IEEE Transactions on Biomedical Circuits and Systems},
  volume = {16},
  number = {3},
  pages = {456--466},
  issn = {1940-9990},
  doi = {10.1109/TBCAS.2022.3182159},
  urldate = {2025-01-29}
}

@inproceedings{haidar2018Design,
  title = {The Design of Fast and Energy-Efficient Linear Solvers: {{On}} the Potential of Half-Precision Arithmetic and Iterative Refinement Techniques},
  booktitle = {Computational Science -- {{ICCS}} 2018},
  author = {Haidar, Azzam and Abdelfattah, Ahmad and Zounon, Mawussi and Wu, Panruo and Pranesh, Srikara and Tomov, Stanimire and Dongarra, Jack},
  editor = {Shi, Yong and Fu, Haohuan and Tian, Yingjie and Krzhizhanovskaya, Valeria V. and Lees, Michael Harold and Dongarra, Jack and Sloot, Peter M. A.},
  year = {2018},
  pages = {586--600},
  publisher = {Springer International Publishing},
  address = {Cham},
  isbn = {978-3-319-93698-7}
}

@article{najafi2024VersaSens,
  title = {{{VersaSens}}: {{An Extendable Multimodal Platform}} for {{Next-Generation Edge-AI Wearables}}},
  shorttitle = {{{VersaSens}}},
  author = {Najafi, Taraneh Aminosharieh and Calero, Jos{\'e} Angel Miranda and Thevenot, J{\'e}r{\^o}me and Duc, Benjamin and Albini, Stefano and Amirshahi, Alireza and Taji, Hossein and Beneyto, Mar{\'i}a Jos{\'e} Belda and Affanni, Antonio and Atienza, David},
  year = {2024},
  month = sep,
  journal = {IEEE Transactions on Circuits and Systems for Artificial Intelligence},
  volume = {1},
  number = {1},
  pages = {83--96},
  issn = {2996-6647},
  doi = {10.1109/TCASAI.2024.3453809},
  urldate = {2025-02-03}
}

@misc{micikevicius2022FP8,
  title = {{{FP8 Formats}} for {{Deep Learning}}},
  author = {Micikevicius, Paulius and Stosic, Dusan and Burgess, Neil and Cornea, Marius and Dubey, Pradeep and Grisenthwaite, Richard and Ha, Sangwon and Heinecke, Alexander and Judd, Patrick and Kamalu, John and Mellempudi, Naveen and Oberman, Stuart and Shoeybi, Mohammad and Siu, Michael and Wu, Hao},
  year = {2022},
  month = sep,
  number = {arXiv:2209.05433},
  eprint = {2209.05433},
  primaryclass = {cs},
  publisher = {arXiv},
  urldate = {2022-09-29},
  archiveprefix = {arXiv},
  langid = {english}
}

\end{document}